\documentclass[aps,prb,showpacs,showkeys,twocolumn]{revtex4}
\usepackage{graphicx}
\usepackage{amsmath}
\usepackage{amssymb}
\usepackage{bm}

\newcommand{\vect}[1] {\mathbf{#1}}   % defining the notation for a vector in
\newcommand{\tens}[1] {\tensor{#1}} % tensor, defined in revsymb.sty
\newcommand{\realpart} {\mathrm{Re}}
\newcommand{\imaginarypart} {\mathrm{Im}}
\newcommand{\xt} {\vect{x},t}
\newcommand{\kw} {\vect{k},\omega}
\newcommand{\dif} {\mathrm{d}}  % differential operator, distinguished from
\newcommand{\evolve} {\frac{\partial}{\partial t}}
\newcommand{\InvariantEnergy} {\biggl(i\evolve-q\varphi(\xt)\biggr)}
\newcommand{\InvariantMomentum}
 {\bigl(-i\nabla-q\vect{A}(\xt)\bigr)}
\newcommand{\Fourier}[1] {\widetilde{#1}}
\newcommand{\FT} {\Fourier{\Sigma_2}}
\newcommand{\GT} {\Fourier{\Sigma_1}}
\newcommand{\psiT} {\Fourier{\psi}}
\newcommand{\Tmat} {\Fourier{\mathcal{T}}}
\newcommand{\DT} {\Fourier{D}}
\newcommand{\AT} {\Fourier{A}}
\newcommand{\measure} { \frac{\dif^dk}{(2\pi)^d}\frac{\dif\omega}{2\pi} }
\newcommand{\Epsilon} {\mathcal{E}}

\allowdisplaybreaks[4]

\begin{document}

\title{Nernst Effect and Anomalous Transport in Cuprates: 
A Preformed-Pair Alternative to the Vortex Scenario}
%\date{\today}
\author{Shina~Tan}
%\email{tansn@uchicago.edu}
\author{K.~Levin}
\affiliation{James Franck Institute and Department of Physics,
  University of Chicago, Chicago, Illinois 60637}

\begin{abstract}
We address those puzzling experiments in 
underdoped high $T_c$ superconductors which have
been associated with normal state ``vortices" and
show these
data can be understood
as deriving from preformed pairs with onset temperature $T^* > T_c$.
For uncorrelated bosons in small magnetic fields, and 
arbitrary $T^*/T_c$,
we present
the exact contribution to \textit{all} transport coefficients. 
In the overdoped regime our results reduce to those
of standard fluctuation theories ($T^*\approx T_c$).
Semi-quantitative agreement
with Nernst, ac and dc conductivity and diamagnetic
measurements is quite reasonable.
\end{abstract}

\pacs{74.40.+k, 74.25.Fy, 74.72.-h}

\keywords{high temperature superconductivity; superconducting fluctuation}

\maketitle

\section{Introduction}

The extent to which the normal state of the high $T_c$ superconductors
is anomalous has long been debated.  The most conclusive
evidence for a break-down of Fermi liquid theory has
appeared relatively recently via a (pseudo)gap in
the fermionic spectrum with onset temperature $T^*$. 
Recent Nernst\cite{Nernst,Ong2} 
and ac conductivity\cite{Corson1999}
experiments have led to some of the most exotic
indications for this pseudogap
which, it is claimed,
appear in the form
of 
``vortices above $T_c$".
More generally (but not universally)
one associates the complex of pseudogap phenomena with
some form of precursor superconductivity.
While normal state
vortices are most closely associated with the well known phase fluctuation
scenario\cite{Emery_Nature},
a primary goal of this paper is to address these same anomalous 
transport
data\cite{Ong2,Corson1999} within the alternative pair fluctuation
scheme\cite{Randeriareview,Chen1998PRL}. In the process, we present a
natural extension of time dependent Ginsburg-Landau
(TDGL) theory and associated transport coefficients which
addresses
higher temperatures $T^*$ well
outside the usual limited range of applicability near $T_c$.  
It necessarily follows that the bosons appear as quantum
rather than classical fields.

The pair fluctuations which we discuss here 
have as a 
natural antecedent the Gaussian 
fluctuations of TDGL.
Indeed, when
pseudogap effects 
are weak ($ T^* \approx T_c$) TDGL theory has been found to provide a
reasonably good
\cite{Larkinreview,Dorsey2,Huse} representation of
the non-critical fluctuation regime in the high temperature superconductors.
In this way the presence of ``pre-formed pairs" has been demonstrated, 
albeit, in the narrow temperature window near $T_c$.
Two important features of the cuprates suggest that
BCS theory is not applicable in that dominant fraction
of the phase diagram associated with the pseudogap phase.
The anomalously short coherence length $\xi$ in the cuprates
combined with the observation of a non-vanishing
excitation gap present at the onset of superconductivity 
has motivated a number of authors\cite{Randeriareview,Chen1998PRL}
to contemplate stronger than
BCS attraction so that the bosonic degrees of freedom
%(and concomitantly, the fermionic pseudogap)
appear at a temperature $T^*$
which may be significantly larger than $T_c$.
Below $T_c$ the counterpart of these pre-formed pairs
appears as non-condensed bosons. The resulting
superconducting state is mid-way between BCS
and Bose Einstein condensation (BEC)\cite{Leggett,NSR,Randeriareview}. 
One can, moreover, establish\cite{JS} that this BCS-BEC crossover approach is 
closely related to (Hartree-approximated) TDGL theory.

Bosonic degrees of freedom
are, in fact, stabilized by going beyond the
usual\cite{Larkinreview,Huse} Gaussian approximation. When their
interactions
are included at the self consistent Hartree level\cite{Hassing,JS} 
a fermionic pseudogap necessarily appears; 
as a result
of this depletion of the density
of states, there are fewer fermions to which bosonic
states can decay.
At the Hartree level the presence of bosons and the existence of a
gap in the fermionic spectrum are two sides of the same coin.
These pair fluctuation approaches differ from other boson-fermion
models\cite{Geshkenbein} of high temperature
superconductors because here the bosons naturally
dissociate into fermions, and similarly fermion pairs
may recombine.
%\cite{Maly,JS}.
%There is, moreover, a microscopic underpinning to TDGL theory which is based
%on a T-matrix, or pair propagator formalism. Thus Gaussian\cite{Larkinreview},
%Hartree\cite{Patton1971}, and more general BCS-BEC schemes\cite{Chen1998PRL,JS,Maly}
%are associated with different levels of approximation to the
%closed set of coupled equations for the T-matrix and fermionic
%propagator.

Hartree approaches to pair fluctuations 
also differ from the widely discussed phase fluctuation
scenario which focuses on fluctuations in the order parameter, and
underlying Mott physics.  In the pair fluctuation approach
the emphasis is on 
the observed small $\xi$. Because it is a mean field theory, this
scheme is not appropriate to the critical
regime and critical exponents here belong to the
Gaussian rather than $3d-~XY$ class.
While Coulombic effects are profound for order parameter
fluctuations in the ordered state, here
they are less important, for precisely the same reasons that they
are omitted in traditional TDGL\cite{Larkinreview}.
As in BCS theory, Coulomb
interactions may also enter
the binding and unbinding of pairs insofar as
they renormalize or even, for non-$s$
wave superconductors\cite{Dongzi},
stabilize the attraction between fermions.

While there are microscopic theories\cite{Larkinreview} which serve to
justify TDGL,
the related diagrammatic formalism 
can become prohibitively complex and
as a consequence, the bosonic contributions to transport properties 
are more readily deduced\cite{Larkinreview} using
TDGL theory directly.
This, then, leads to the present focus on establishing
a quantum extension of TDGL which is thereby amenable to
detailed transport studies, at higher temperatures $T^*$.
To proceed, we
construct a simplified model of charged bosons subject to quantum
dissipation, which has TDGL (with a non-BCS parameter set)
as its special limit.
Pair fluctuations involve the continuous dissociation and recombination
of fermion pairs.
To simulate this behavior
we follow Caldeira and Leggett
and treat the fermions as
reservoir harmonic oscillators\cite{Caldeira,Guinea}
coupled to the bosons.
Coulomb and hard core interactions between bosons are treated
at the same level as in Hartree-approximated TDGL.

We end by summarizing the key experiments we will address here.
We focus on those experiments where the bosonic contributions
dominate over their fermionic counterparts. It is convenient to
define
the electrical current $\vect{J^0} =
 \sigma \vect{E} + \alpha \vect(-\nabla T)$.
In contrast to the behavior in a Fermi liquid,
a sizeable Nernst signal $\nu $ (reflecting a combination
of components of
the $\alpha$ and $\sigma$ tensors\cite{Ong2}) 
% $\approx  \alpha_{xy}/\sigma_{xx}(1/B)$???
appears at an onset temperature appreciably above
$T_c$, called $T_{\nu}^*$; this
temperature is loosely
associated with $T^*$.
In the
insulating phase, the thermoelectric response function $\alpha_{xy}$, 
as well as $\nu$ vanish at $T=0$. 
Precursor effects are similarly observed\cite{Corson1999} in the
imaginary component, $\sigma_2$, of $\sigma$
at $100~\mathrm{GHz}$, 
which set in below $T_{\sigma}^*$, but appreciably above $T_c$.
Thus far, these onset temperatures are significantly below
their counterparts for $\alpha_{xy}$.
Moreover, for the range of frequencies
measured, these complex conductivity
data can be fit\cite{Corson1999} to a rescaled Kosterlitz-Thouless
form which allows extraction of a phase correlation time $\tau'(T)$.
In view of the above $\sigma$ and $\alpha$ experiments,
it has seemed rather mysterious that
the diamagnetic magnetization\cite{Ong2}, has a relatively 
non-existent normal state precursor.

Finally, the dc resistivity
has been measured somewhat
systematically\cite{Watanabe1997PRL,Leridon2001PRL,Curras}
in the underdoped regime. There is
evidence for an onset temperature $T_{\rho}^*$ of roughly
the same order of magnitude as in the other transport
experiments. At this temperature an
enhancement of the conductivity 
is evident.  However, the magnitude of the deviation from the fermionic
contribution is not noticeably large except in
the immediate vicinity of $T_c$.  Analysis of these data do,
however, depend on assigning a particular temperature dependence
to this background fermionic contribution.

\section{The Uncorrelated Boson Model}

\subsection{Revisiting Time Dependent Ginsburg Landau Theory}

The time dependent Ginsberg-Landau equation (TDGL) 
\begin{multline}\label{eq:TDGL}
\gamma\InvariantEnergy\psi(\xt)=\frac{\InvariantMomentum^2}{2m}\psi(\xt)\\
+a_0(T)\psi(\xt)+a'\lvert\psi(\xt)\rvert^2\psi(\xt)+D(\xt)
\end{multline}
is a natural starting point for characterizing the dynamics of bosons.
This equation describes the fluctuational regime 
of conventional superconductors
in the vicinity of $T_c$,
where the bosonic degrees of freedom
are represented by the wavefunction $\psi(\xt)$.
Here $\lvert q\rvert=2e$ represents the bosonic charge,
($\varphi$, $\vect{A}$) is the electromagnetic potential 
%($d+1$)-vector,
which determines the EM fields via the usual formulae
$\vect{E}(\xt)=-\nabla \varphi(\xt)-\evolve\vect{A}(\xt)$, and 
$\vect{B}(\xt)=\nabla\times\vect{A}(\xt)$.
The units adopted in this paper correspond to SI, with
$\hbar = \epsilon_0 = k_B = 1$.
The dynamics is importantly controlled by 
$D(\xt)$ which is a white noise function satisfying
\begin{equation}\label{eq:TDGL_noise}
\bigl\langle D^*(\vect{x'}t')D(\xt)\bigr\rangle=2\gamma_2T\delta(t-t')\delta(\vect{x}-\vect{x'}),
\end{equation}
Generally we contemplate
complex $\gamma$ with $\gamma_2 \equiv\imaginarypart\gamma$
and $\gamma_1\equiv\realpart\gamma$.
%, $\gamma_2\equiv\imaginarypart\gamma$.

In this paper we focus on a description in which bosonic degrees of
freedom may be treated as uncorrelated. For the TDGL case this 
corresponds to applying a  
Hartree approximation to 
Eq.~\eqref{eq:TDGL} so that,
\begin{multline}\label{eq:HartreeTDGL}
\gamma\InvariantEnergy\psi(\xt)=\frac{\InvariantMomentum^2}{2m}\psi(\xt)\\
+a(T)\psi(\xt)+D(\xt),
\end{multline}
%is often used by people\cite{Huse}. 
Here $a(T)$ represents the 
(absolute value of the) bosonic chemical potential which vanishes at the 
superconducting
transition temperature $T_c$, which is to be distinguished from
$T_c^\text{mf}$ at which $a_0 = 0$.
Frequently one ignores the quartic term altogether, as in
strict Gaussian fluctuation theories\cite{Huse}. An
interpretation as well as justification for TDGL 
can be provided by microscopic T-matrix approaches, based
on the well known Aslamazov-Larkin diagrams\cite{Larkinreview}.
With the inclusion of Hartree effects\cite{Hassing}
the fluctuations have characteristic Gaussian exponents 
while Hartree self consistency contributions lead to a slightly
modified T-matrix scheme\cite{Patton1971}.

At the more macroscopic level of Eq.(\ref{eq:TDGL}) the fermions
in a superconductor are irrelevant. Nevertheless, microscopic
theory makes it clear that the bosonic degrees of freedom
correspond to fermion pairs; moreover, it is the fermions
which are ultimately responsible for the complex noise
parameter $\gamma$.
The self consistent
Hartree approximation introduces
a fermionic exitation gap\cite{Hassing} (or ``pseudogap")
which is present at the onset of superconductivity.  This
depletion in the density of states
is responsible for the fact that the transition temperature
(contained in $a(T)$)
is lowered, relative to $T_c^\text{mf}$.

Despite its significant success in describing conventional
superconducting fluctuations, TDGL has known limitations\cite{words}.
%SHINA can you put this in above footnote:
% First, the white-noise leads to some divergent results,
%including $\langle\psi^*\psi\rangle$ and some transport 
%coefficients such as the thermal conductivity.
In the context of understanding high $T_c$ superconductors,
one of the most serious of these
is the necessity of introducing artificial cut-offs
in the fluctuational spectrum\cite{Silva}, often to
depress the fluctuational contributions to transport. 
Other extensions of TDGL have been proposed which involve introducing a 
modification\cite{Geshkenbein}
to the BCS-derived parameter set of Eq.(\ref{eq:TDGL}),
although for some experiments\cite{Huse} strict BCS theory
appears to work quite well.

What is most perplexing about high $T_c$ superconductors is
the appearance of ``pseudogap" effects
with onset temperature $T^*$. As $T^*$ progressively increases
away from $T_c$,
a strict BCS approach to fluctuation-based calculations
of transport appears to be invalidated\cite{Huse}.
\textit{It is the premise of the present paper that the 
precursor superconductivity of the pseudogap
phase evolves continuously from
the conventional fluctuation behavior} seen, for example, in overdoped high
$T_c$ samples.  While
one does not expect TDGL to hold for $T$ significantly larger than $T_c$,
our goal here is
to propose a natural extension of this theory appropriate for
$T$ well above $T_c$.
In this regime, the relaxation time of the bosons becomes comparable to or 
smaller than $\hbar/k_BT$; thus,
the classical fluctuation-dissipation in TDGL should be replaced by
a suitable quantum counterpart. 
In this way we treat
$\psi$ in Eq.(\ref{eq:TDGL})
as a field theoretic operator representing the annihilation
of bosons. 
One of the most important parameters in this
extended TDGL theory is $T^*$ which will enter into
the bosonic chemical potential. 
%(See Eq.(\ref{eq:quantum_mu_pair})below).
 This is the temperature
at which the number of bosons vanishes. One may view $T^*$ alternatively
as the
onset of the fermionic pseudogap. These two viewpoints are two
sides of the same coin, since bosons disappear or dissociate when fermions
are no longer bound.
%It should not be assumed that this is a sharp transition temperature
%but rather a continuous crossover temperature scale.

\subsection{Pre-Formed Pair Model: Extended TDGL }

To gain a deeper understanding of the essence of TDGL
and of its prior success away from the pseudogap 
(or underdoped) regime\cite{Huse},
we thus study a Hamiltonian describing bosons on a $d$-dimensional lattice coupled
to a quantum reservoir. Our treatment of the reservoir has
strong similarities to the approach of Caldeira and Leggett\cite{Caldeira}.
We consider  
\begin{multline} \label{eq:H}
H=\sum_{\vect{u}\vect{x}}\varepsilon_{\vect u\vect x}\psi_{\vect{x}}
^\dagger(t)\exp
\bigl(-iqC_{\vect u\vect x}(t)\bigr)\psi_{\vect{x}+\vect{u}}(t)\\
+\sum_{\vect x}q\varphi\psi^\dagger\psi
+\sum_{i\vect x}\Bigl\{(a_i+q\varphi)w_i^\dagger w_i+\eta_i\psi^\dagger w_i
+\eta_i^*w_i^\dagger\psi\Bigr\}\\
+\sum_{i\vect x}\Bigl\{(b_i-q\varphi)v_i^\dagger v_i+\zeta_i\psi^\dagger v_i
^\dagger
+\zeta_i^*v_i\psi\Bigr\}.
\end{multline}
Here $\psi_{\vect{x}}(t)$ is the boson annihilation operator
at lattice site $\vect{x}$ and time $t$,
%($\varphi$, $\vect{A}$) is the electromagnetic potential,
%$\hbar = \epsilon_0=k_B=1$ and 
%Moreover,
%$\varepsilon_{\vect u\vect x}$ 
%is the hopping matrix element. 
Annihilation operators for the reservoir,
$w_i$ and $v_i$ (with infinitesimal coupling constants $\eta_i$
and $\zeta_i$), are associated with
positive and negative frequencies respectively, although 
the energies $a_i$'s and $b_i$'s are all positive.
That two sets of reservoir
operators are necessary will become clear
later when we compare with standard TDGL.

%The coupling coefficients $\eta_i$ and $\zeta_i$ are infinitesimal\cite{Caldeira},
%for each $i$; nevertheless, the total contribution from the entire set
%is arbitrarily large.
Here $\varepsilon_{\vect u\vect x}$ is the hopping matrix element of the bosons,
to be distinguished from its electronic counterpart, 
and
$C_{\vect u\vect x}(t)\equiv \int_0^1\dif su_aA_a(\vect x+s\vect u,t)$. 
%is an electromagnetic phase factor 
%evaluated on the straight line path connecting two lattice sites.
Surface effects appear via
the $\vect x$ dependence of  
$\varepsilon_{\vect 0\vect x}$   
%($\vect u\neq \vect 0$) will be independent of $\vect x$; similarly,
%$\varepsilon_{\vect 0\vect x}$ will be so inside the bulk volume.
%Surface effects can be modeled by the $\vect x$ dependence of
% $\varepsilon_{\vect 0\vect x}$,
%which gradually increases to $+\infty$ toward the boundary, so that the 
%boson density decreases
%to zero.
which also contains the
Hartree interaction between bosons.
%, which leads to a shift in their
%chemical potential, $\mu_\mathrm{pair}$.
%The equations of motion of these 
%bosonic operators are precisely gauge invariant.
%Finally, we
%consider anisotropic, layered
%superconductors\cite{Lawrence} here, which are associated with a general
%lattice bandstructure for the bosons.

The equations of motion of the system are given by 
\begin{subequations}\label{eq:motion}\begin{align}
\InvariantEnergy\psi_{\vect x}(t)&=\sum_{\vect u}\varepsilon_{\vect u\vect x}
\exp\bigl(-iqC_{\vect u\vect x}(t)\bigr)\psi_{\vect x+\vect u}(t)\notag\\
+\sum_i&\eta_{i\vect x}w_{i\vect x}(t)+\sum_i\zeta_{i\vect x}v_{i\vect x}^\dagger(t),
\label{eq:motiona}\\
\InvariantEnergy w_{i\vect x}(t)&=a_{i\vect x}w_{i\vect x}(t)
+\eta_{i\vect x}^*\psi_{\vect x}(t),\label{eq:motionb}\\
\InvariantEnergy v_{i\vect x}^\dagger(t)&=
-b_{i\vect x}v_{i\vect x}^\dagger(t)-\zeta_{i\vect x}^*\psi_{\vect x}(t).
\label{eq:motionc}
\end{align}\end{subequations}
This model is manifestly gauge invariant.

We solve
Eqs.~\eqref{eq:motionb} and \eqref{eq:motionc} in the temporal gauge 
($\varphi(\xt)\equiv 0$) to 
express $w_i$ and $v_i^\dagger$ in terms of 
both
$\psi_\vect{x}(t)$ and their values at an initial time $t_0\rightarrow-\infty$. Substituting 
into Eq.~\eqref{eq:motiona}, we find
\begin{multline}\label{eq:psimotion}
i\evolve\psi_{\vect x}(t)=\sum_{\vect u}\varepsilon_{\vect u\vect x}\exp
\bigl(-iqC_{\vect u\vect x}\bigr)\psi_{\vect x+\vect u}(t)\\
-i\int_{t_0}^{t}\dif t'\Sigma_2\bigl(t-t',T_{\vect x}\bigr)\psi_{\vect x}(t')+D_{\vect x}(t),
\end{multline}
where $T_{\vect x}$ is the local temperature
and $\Sigma_2(t)$ the Fourier transform of
\begin{equation}\label{eq:Sigma2}
\FT(\omega)\equiv 2\pi\sum_i\lvert\eta_i\rvert^2\delta(\omega-a_i)
    -2\pi\sum_i\lvert\zeta_i\rvert^2\delta(\omega+b_i),
\end{equation}
It is reasonable to
assume that this self energy, like any other, is smooth and that it
vanishes as $\omega\rightarrow\pm\infty$. Here
\begin{multline}\label{eq:quantumnoise}
D_{\vect x}(t)\equiv \sum_i\exp\bigl(-ia_{i\vect x}(t-t_0)\bigr)\eta_{i\vect x}
w_{i\vect x}(t_0)\\
+\sum_i\exp\bigl(+ib_{i\vect x}(t-t_0)\bigr)\zeta_{i\vect x}v_{i\vect x}^\dagger(t_0)
\end{multline}
is a function which represents a generalized or quantum noise.
The coupling between the $\psi$ field and each reservoir field
is infinitesimal, so that
the reservoir satisfies ideal Bose statistics and
\begin{multline}\label{eq:quantumnoiseamplitude}
\bigl\langle D_{\vect x'}(t')^\dagger D_{\vect x}(t)\bigr\rangle
=\delta_{\vect x\vect x'}\int\frac{\dif\omega}{2\pi}\FT(\omega,T_{\vect x})
b(\omega,T_{\vect x})\\
\cdot\exp\bigl(-i\omega(t-t')\bigr),
\end{multline}
with
$b(\omega,T)\equiv 1/\bigl(\exp(\omega/T)-1\bigr)$ the Bose function.
The physics of the bosons, transport, magnetization, specific heat, density, etc,
are governed by Eqs.~\eqref{eq:psimotion} and \eqref{eq:quantumnoiseamplitude}.

The reservoir parameters $a_i$, $b_i$, $\eta_i$ and $\zeta_i$
are all subsumed into the boson self energy $\FT(\omega)$. From
this point forward we ignore these quantities in favor
of the
boson self energy.
Moreover, $\FT(\omega)$ depends only on $\omega$ for our localized
reservoir. This simplification is supported principally by the
fact that this model captures the key physics found in 
microscopic schemes\cite{Maly,Randeriareview}, 
where it is sufficient to consider
just the leading order
$\vect k$ dependences.
Differences between this
previous strong coupling T-matrix approach\cite{Maly,Randeriareview}
and the present phenomenological
boson model, are to be associated with the fact that the former
scheme assumes that the T-matrix
has fermionic constituents. These lead to a fine $\omega$
structure arising from the fermionic pseudogap\cite{Maly},
as well as to different high $\omega$ asymptotics. The ensuing simplification
of boson physics, however, makes our model far more tractable.

In the spatially uniform case, the Fourier transform of 
Eq.~\eqref{eq:psimotion} takes a
simple form:
\begin{equation}\label{eq:psimotion_uniform}
\Bigl(\omega-\varepsilon(\vect k)-\GT(\omega)+\frac{i}{2}\FT(\omega)\Bigr)\psiT(\kw)
=\DT(\kw),
\end{equation}
with $\varepsilon(\vect k)\equiv\sum_{\vect u}
\varepsilon_{\vect u}\exp(i\vect k\hspace{-4pt}\cdot\hspace{-4pt}\vect u)$,
and 
\begin{equation}
\GT(\omega)\equiv
P\int\frac{\dif\omega'}{2\pi}\frac{1}{\omega-\omega'}\FT(\omega').
\end{equation}

We are now in a position to clarify the relationship
between this theory and TDGL. When $T$ is close to $T_c$,
the bosonic relaxation rate $\tau^{-1}$
is considerably smaller than $T$ (typically of the order of tens
or hundreds of Kelvin). In this
regime the dynamics is dominated by low frequencies
$\lvert\omega\rvert\sim \tau^{-1} \ll T$, and the Bose function is well
approximated by
$b(\omega,T)\approx T/\omega$.  We presume that the
high energy cut-off scale $\Omega$ in
$\FT(\omega)$, which is associated with the
reservoir, is of the order
of typical electronic energies, thus thousands of K.
Then, at low frequencies, the bosonic
self energy is given by the linear functions
\begin{equation}\label{eq:Glinear}
\GT(\omega)-\GT(0)\approx(1-\gamma_1)\omega
\end{equation}
and
\begin{equation}\label{eq:Flinear}
\FT (\omega) \approx 2 \gamma_2 \omega,
\end{equation}
where from Eq.~\eqref{eq:Sigma2},
$\FT(\omega)$ vanishes at zero frequency.

With the above approximations 
the quantum noise correlation function
Eq.~\eqref{eq:quantumnoiseamplitude} is reduced to its classical limit, 
given by
Eq.~\eqref{eq:TDGL_noise}. In the
same way Eq.~\eqref{eq:psimotion_uniform} is reduced to
Eq.~\eqref{eq:HartreeTDGL}, with the important parameter 
\begin{equation}\label{eq:quantum_mu_pair}
-\mu_\mathrm{pair}\equiv a(T)=\min_{\vect k}\varepsilon(\vect k)+\GT(0).
\end{equation}
We may now see that
the two sets of reservoir fields, $w_i$ and $v_i$, are
associated with opposite charges,
$q$ and $-q$; they
contribute to the positive and negative frequency regimes of
$\FT(\omega)$, respectively. If either
of these two were omitted, $\FT(\omega)$
would vanish on one side of the origin; thus, the slope of $\FT(\omega)$ 
would be
discontinuous across zero frequency, and one  
would never arrive at the TDGL limit,
no matter how small the frequency.
\textit{In summary, the boson model presented in this paper is a natural
quantum extension of TDGL. 
Conversely, TDGL is the low frequency limit of our quantum
boson model}.

In preparation for computing
transport coefficients, we introduce
the bosonic correlation function, which
can be derived from Eqs.~\eqref{eq:psimotion_uniform}
and \eqref{eq:quantumnoiseamplitude}. This is given by
\begin{equation}\label{eq:field_correlation_function}
\langle\psiT^\dagger(\vect k'\omega')\psiT(\kw)\rangle
=\AT(\kw)b(\omega)(2\pi)^{d+1}\delta(\vect k-\vect k')\delta(\omega-\omega')
\end{equation}
%in the spatially uniform case, 
where $\AT(\kw)=\realpart 2i\Tmat(\kw)$
is the boson spectral function,
and 
\begin{equation}\label{eq:vertex}
\Tmat(\kw) \equiv \Bigl(\omega-\varepsilon(\vect k)-\GT(\omega)
+\frac{i}{2}\FT(\omega)\Bigr)^{-1}
\end{equation}
is the boson propagator or ``T-matrix". The average number of bosons per lattice site is
\begin{equation}\label{eq:density}
n\equiv \langle\psi^\dagger\psi\rangle=v\int\measure
\AT(\kw)b(\omega),
\end{equation}
where $v$ is the cell volume associated with each lattice site.

\subsection{Transport Coefficients}

The electric current $\vect J^0$ and heat current $\vect J^1$ follow from
charge and energy conservation:
\begin{multline}\label{eq:J}
J^n_a(\xt)=\frac{q^{1-n}}{2v}\biggl\{\InvariantEnergy^n\psi(\xt)\biggr\}
^\dagger\\
\cdot\sum_{\vect u}iu^a\varepsilon_{\vect u\vect x}
\exp\bigl(-iqC_{\vect u\vect x}(t)\bigr)
\psi_{\vect x+\vect u}(t)+\text{h.c.}.
\end{multline}
When a magnetic field $B_{ab}\equiv\frac{\partial}{\partial x_a}A_b
-\frac{\partial}{\partial x_b}A_a$ is applied to the system, surface 
electric and heat currents appear in a thin shell around its
boundary.

To derive these surface magnetizations we proceed as follows. 
We confine bosons within the sample boundaries via
a spatially
dependent $\varepsilon_{\vect{0}\vect{x}}$ which approaches the
bulk value well away from the surface and $+\infty$ on the
boundary. We presume that
the spatial gradient of $\varepsilon_{\vect 0\vect x }$
is small, as is the applied magnetic field so that we may calculate
%Eq.~\eqref{eq:psimotion} perturbatively. The
the electric and heat currents to leading order in these
quantities. 
Integrating the current in the normal direction of the surface,
from the boundary to deep within the sample, we obtain
%get Eq.~\eqref{eq:M}.
%In this derivation, we employ the assumption that the spatial gradient of
%$\varepsilon_{\vect 0\vect x}$ is small. But we believe that the result,
%Eq.~\eqref{eq:M}, is independent of it.
\begin{equation}\label{eq:M}
M^n_{ab}=\frac{q^{2-n}}{6}B_{cd}\int\measure\Delta_{abcd}\realpart i\Tmat(\kw)
^2\omega^nb(\omega),
\end{equation}
where $M^0_{ab}$ is the usual magnetization, $M^1_{ab}$ is 
the thermal analogue\cite{footnote2} 
of $M^0_{ab}$,
$\Delta_{abcd}(\vect k)\equiv\frac{1}{2}\bigl(v_{ac}(\vect k)v_{bd}(\vect k)$
$-v_{ad}(\vect k)v_{bc}(\vect k)\bigr)$, and $v_{ab}(\vect k)\equiv
\frac{\partial^2}{\partial k_a
\partial k_b}\varepsilon(\vect k)$ is the inverse mass tensor
of the bosons, whose group velocity is given by
$v_a(\vect k)\equiv\frac{\partial}{\partial k_a}\varepsilon(\vect k)$.

When an external magnetic field is present, the bulk volume
current $\vect{J}^n$ must be combined with surface contributions
so that the net
``transport" currents 
are given by\cite{Huse}:
\begin{equation}\label{eq:Jtr}
J^n_{\text{(tr)}a}=\langle J^n_a\rangle
+\frac{\partial M^n_{ab}}{\partial T}E^1_b+\delta_{n,1}M^0_{ab}E^0_b.
\end{equation}

For small, but constant $B_{ab}$,
 electric field $E^0_a\equiv E_a$, and
negative thermal gradient $E^1_a\equiv-\frac{\partial}{\partial x_a} T$,
we obtain the linearized response functions
$J^n_{\text{(tr)}a}=\sum_{n'=0}^1\sum_bL^{nn'}_{ab}E^{n'}_b$, 
and the DC transport
coefficients $L^{nn'}_{ab}$ can be compactly written as
\begin{multline}\label{eq:L}
L^{nn'}_{ab}=\frac{q^{2-n-n'}}{2T^{n'}}\int\measure v_av_b\AT^2\omega^{n+n'}
b^{(1)}\\
+\frac{q^{3-n-n'}B_{cd}}{6T^{n'}}\int\measure v_av_cv_{bd}\AT^3\omega^{n+n'}
b^{(1)},
\end{multline}
where $b^{(1)}(\omega)\equiv -\frac{\partial b(\omega)}{\partial\omega}$
arises from the $\omega =0$ limit, in which the boson absorbs or emits
an infinitesimal amount of energy in the
process of making a transition to an adjacent energy level.
 $L^{00}_{ab}=\sigma_{ab}$ is the isothermal electric conductivity,
 $L^{11}_{ab}$ is the isoelectropotential
thermal conductivity, and $L^{01}_{ab}=\alpha_{ab}$ and $L^{10}_{ab}$
 are off-diagonal coefficients.
Eq.~\eqref{eq:L} satisfies 
the Onsager relation\cite{Huse} $T^{n'}L^{nn'}_{ab}(\tens B)=
T^{n}L^{n'n}_{ba}(-\tens B)$ (no summation), as a consequence
of our inclusion of the
surface terms\cite{Huse}.
Surface effects enter into $L^{01}_{ab}$,
$L^{10}_{ab}$ and $L^{11}_{ab}$, but 
cancel in $L^{00}_{ab}$.

Finally, we deduce the
boson contribution to the complex ac conductivity and ac Hall conductivity
\begin{multline}\label{eq:ACconductivity}
\Fourier{\sigma}_{ab}(\omega)=\frac{iq^2}{\omega}\int\frac{\dif^dk'}{(2\pi)^d}
\frac{\dif\omega'}{2\pi}v_a(\vect{k'})v_b(\vect{k'})
\AT(\vect{k'}\omega')b(\omega')\\
\cdot
\bigl(\Tmat(\vect{k'},\omega'+\omega)
+\Tmat^*(\vect{k'},\omega'-\omega)
-\Tmat(\vect{k'}\omega')-\Tmat^*(\vect{k'}\omega')\bigr)\\
+\frac{iq^3}{2\omega}B_{cd}\int\frac{\dif^dk'}{(2\pi)^d}
\frac{\dif\omega'}{2\pi}
v_a(\vect k')v_c(\vect{k'})v_{bd}(\vect{k'})\AT(\vect{k'}\omega')b(\omega')\\
\cdot
\bigl(i\Tmat(\vect{k'},\omega'+\omega)
-i\Tmat^*(\vect{k'},\omega'-\omega)\bigr)\\
\cdot
\bigl(\Tmat(\vect{k'},\omega'+\omega)
+\Tmat^*(\vect{k'},\omega'-\omega)
-\Tmat(\vect{k'}\omega')-\Tmat^*(\vect{k'}\omega')\bigr).
\end{multline}
It can be verified that $\Fourier{\sigma}_{ab}(\omega\rightarrow 0)
=L^{00}_{ab}$ and that 
Eq.~\eqref{eq:ACconductivity} satisfies the f-sum rule.
%\begin{equation}\label{eq:sumrule}
%\int_{0}^{+\infty}\dif\omega\realpart\Fourier{\sigma}_{ab}(\omega)
%=\frac{\pi q^2}{2}\int\measure v_{ab}(\vect{k})\AT(\kw)b(\omega).
%\end{equation}
%
%Our numerical evaluation 
%of these transport coefficients 
%will make use of the fact that the 
%we note that because $\AT$ is dependent of $\vect k$ via
% $\varepsilon(\vect k)$, 
%integrals over momentum in Eqs.~\eqref{eq:L} and \eqref{eq:ACconductivity}
%can be reduced to integrals over a single energy variable
%by introducing generalized density of states functions. 
%We presume that 
%$T^*/T_c$ (which is roughly fitted
%to the phase diagram) is the principle input parameter rather than
%hole concentration $x$, or coupling constant $g$.
%Essential to this theory is the
%temperature dependence of
%the boson
%chemical potential $\mu_\mathrm{pair}(T)$
%which, is chosen so that the
%%=\max_{\vect k}\GammaT(\vect k,0)$. 
%In 
%the TDGL model 
%quantity is proportional to $T- T_c$. 
%boson number is zero above $T^*$.
%This should be contrasted with
%the TDGL result $\mu_\mathrm{pair}\propto (T-T_c)$ which holds
%only near Bose condensation\cite{footnote1}, and generally tends
%to overestimate
%bosonic effects away from $T_c$.

\section{Phenomenology}

In this section we
arrive at a simple and generic phenomenology
describing the pre-formed pairs or bosons
of our extended TDGL model,
in the context of hole-doped cuprates. Our goal in the
next section is to use this phenomenology to address
transport data
at a semi-quantitative level with as few fitting
parameters as possible. It should be noted that this
phenomenology is generally compatible with previously derived
T-matrix based approaches\cite{Chen1998PRL}. 
The most important component of this phenomenological
discussion lies in our introduction of the parameter
$T^*$, which has no natural counterpart in TDGL
approaches. This is the temperature at which the number
of bosons vanishes. This same temperature is reflected
in the fermionic spectrum as that at which the excitation
(pseudo)gap vanishes, so that fermions are no longer bound
into bosons. The remaining parameter choices discussed below
are reasonably straightforward, and represent natural
extensions based on TDGL theory.

There are three factors
which govern the dynamics of 
our bosons: the dispersion function $\varepsilon(\vect k)$,
the self energy $\widetilde{\Sigma}(\omega)$, and the chemical potential $\mu_\mathrm{pair}$.
We begin with $\varepsilon(\vect k)$, which is
given by $\Epsilon_a\bigl(1-\cos(k_as_a)\bigr)
+\Epsilon_b\bigl(1-\cos(k_bs_b)\bigr)+\Epsilon_c\bigl(1-\cos(k_cs_c)\bigr)+
\text{const.}$,
presuming only nearest neighbor hopping.
Here $s_c\equiv s$ is the interlayer
spacing, while $s_a$ and $s_b$ are in-plane lattice constants. The 
characteristic boson bandwidths
($\Epsilon_{a,b}$) satisfy
$\Epsilon_a\approx\Epsilon_b$ of the order of a few thousand K\cite{Footnote_widetilde}. 
By contrast, $\Epsilon_c$
is on the order 10K for the least anisotropic cuprates (e.g., YBCO)
and roughly two orders of magnitude
smaller for the most anisotropic systems (e.g, Bi2212).
In this way the dispersion function may be further
simplified to yield
$k_{ab}^2/2m_{a}+\Epsilon_c\bigl(1-\cos(k_cs)\bigr)+\text{const.}$,
the well known Lawrence Doniach dispersion.

Next we turn to $\widetilde{\Sigma}(\omega)$,
which is expected to be given by its TDGL form.
Eqs.~\eqref{eq:Glinear} and \eqref{eq:Flinear}, for
frequencies much lower than a characteristic
cutoff energy $\Omega$, introduced
earlier.  
Moreover, in the fermionic regime, far from the true bosonic limit,
(as appears to be the case for the
cuprates),
$\gamma_1\ll\gamma_2$. Our results are, thus, rather insensitive
to $\gamma_1$ and for simplicity we set it to zero. \cite{Footnote_gamma1}
% If we are to study the Hall conductivity etc, the first order
%term in $\gamma_1$ is dominant, since the zeroth order term vanishes. 
%See Ref.\onlinecite{Huse}.

To obtain an estimate of
$\Omega$, we use our model to calculate
the average number of bosons per lattice site $n(T)$. At $T_c$ we obtain
the simple result\cite{Footnote_n_Tc}:
\begin{equation}\label{eq:n_Tc}
n(T_c)\approx\frac{\widetilde{\Omega}}
{4\pi\widetilde{\Epsilon}_a}
+\frac{T_c}{2\pi\Epsilon_a}\ln\frac{2T_c}{\widetilde{\Epsilon}_c},
\end{equation}
where, for general energy scales $E$ we define 
$\widetilde{E}\equiv E/\gamma_2$. 
It is reasonable to assume
that well into the underdoped regime $n(T_c)$ falls somewhere inside
the range
0.01-0.5 electron pairs per lattice site.
Since $T_c$ is considerably
smaller than $\widetilde{\Epsilon}_{a}$,
the first term on the right side of Eq.~\eqref{eq:n_Tc} dominates, and 
we deduce that $\Omega$ is
of the same order of magnitude as $\Epsilon_a$, thus
thousands of K.
The above analysis indicates that
the simple linearized expansion of the boson
self energy, given by Eqs.~\eqref{eq:Glinear} and \eqref{eq:Flinear},
is a reasonably
good approximation over the range of relevant frequencies we
consider here.

Finally we address the quantity
$\mu_\text{pair}(T)$ which depends on the important temperature $T^*$ which has
no natural counterpart in TDGL theory.  This is the
temperature where $\mu_\text{pair}$ diverges. At this temperature
the Bose degrees of freedom vanish.  Concomitantly, at $T^*$ the fermionic
excitation gap disappears, although, presumably this is
a crossover scale rather than a sharp transition.  It has
been argued that TDGL approaches overestimate fluctuation
effects so that short wavelength cut-offs in the fluctuational
spectrum have to be introduced, most recently to
address the paraconductivity\cite{Silva}.  In the boson
model introduced here, the non-TDGL temperature
dependence in the pair chemical potential,
$\mu_\text{pair}$, removes
the necessity for introducing an ad hoc cut-off.

We may estimate the
magnitude of $\mu_\text{pair}$ at intermediate temperatures
between $T_c$ and $T^*$, by using the boson density $n(T)$. 
T-matrix based theories suggest\cite{Chen1998PRL}, that the electronic pseudogap
scales with the number of bosons above $T_c$.  A key assumption
of our approach is that the magnitude of the pseudogap
decreases by
an appreciable fraction at intermediate temperatures
between $T_c$ and $T^*$. This assumption is reasonably compatible
with experiments such as ARPES, but it has not been
conclusively established at this time.
We expect that the number of bosons
evolves in a similar fashion. 

Within our
theoretical model,
%$\mu_\text{pair}$ has to be on the order of $\Omega$ to supress $n$ 
%significantly 
%relative to $n(T_c)$. 
we require that $-\mu_\text{pair}\sim\Omega$ in order to supress
$n(T)$ significantly relative to $n(T_c)$,
for $T$
between $T_c$ and $T^*$\cite{Footnote_transport}.
TDGL expressions for $\mu_\text{pair}$
($\mu_\text{pair}\propto (T-T_c)$)
will not lead to a sufficiently rapid decline in the boson
number.
It follows that there must be an additional term in
the pair chemical potential
%$\mu_\text{pair}$ 
which
is negligible in the vicinity of $T_c$, but
which rapidly increases at higher temperatures, and diverges
at $T^*$. 
A form generally compatible with the above analysis, as well as with
microscopic T-matrix based schemes\cite{Chen1998PRL},
can be written as
\begin{equation}\label{eq:a_1}
\tau^{-1}(T)\equiv-\frac{\mu_\text{pair}}{\gamma_2}
=\frac{8}{\pi\eta}(T-T_c)+c\frac{(T-T_c)^3}{(T^*-T)^2},
\end{equation}
where we have introduced a quantity
$\eta$ which represents the ratio of the boson coherence 
time $\tau$
%\equiv 1/\widetilde{a}(T)$
to its BCS value, $\pi/\bigl(8(T-T_c)\bigr)$, for $T\sim T_c$. 
Throughout, we take $c\sim 15$, 
consistent with our order of magnitude estimates of $\mu_\text{pair}$.
It should be stressed that
there is nothing in this and the following sections which depends
on the specific details of this functional form, provided
the non-TDGL contribution (or second term) 
diverges at $T^*$
and vanishes sufficiently rapidly as $T$ approaches $T_c$.

Both ac and dc conductivity data indicate that $\eta$ is close to $1$ for optimally doped
cuprates\cite{Balestrino1992PRB} but it
becomes significantly larger than $1$ as the material is progressively
underdoped\cite{Corson1999}. In this way
the pairs live longer in the vicinity of $T_c$ than expected from BCS theory.
Quantitative analysis of
the ac conductivity data (presented in the next section) shows that $\eta$ can be of the order
10-20 for a typical underdoped Bi2212 sample.
\textit{We take $\eta$ as the only essential fitted parameter
in all of our numerical analysis}\cite{Footnote_eta}. Other parameter 
choices\cite{Footnote_other} 
throughout this paper are for
illustrative purposes only.

We end this section by noting that there is an additional
quantum statistical
effect which acts to further suppress boson transport.
This is associated with the fact that
the factor $-b'(\omega)$ is smaller than its TDGL counterpart $T/\omega^2$.
This effect is appreciable when $\tau\sim1/T$, and greatly amplified
when $\tau\ll1/T$,
thereby further suppressing bosonic
contributions to (DC) transport.

\section{Numerics and Qualitative Comparison with Experiment}
\subsection{Nernst Effect and Magnetization}

\begin{figure}[!thb]
\centerline{\includegraphics{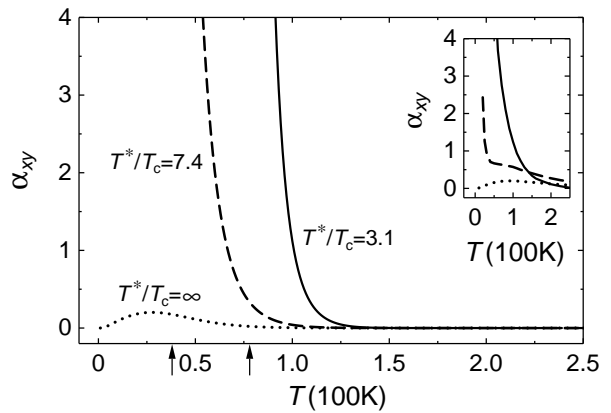}}
\caption{Theoretical curves for the normal state $\alpha_{xy}$
with variable $T^*/T_c$; experimental
counterparts\cite{Ong3} in inset for 
$\mathrm{La}_{2-x}\mathrm{Sr}_x\mathrm{Cu}\mathrm{O}_4$ at $x=0.03$, $0.07$,
and $0.12$. Both the theoretical and experimental values of $\alpha_{xy}$ are
 normalized by 
$0.02 (B/\mathrm{T})\mathrm{V}/(\mathrm{K\Omega m})$, a number
obtained earlier\cite{Huse}  
at $2T_c$ for an optimally doped sample.
\label{fig:alpha}}
\end{figure}

In Fig.~\ref{fig:alpha} we plot the transverse thermoelectric
 coefficient $\alpha_{xy}$ vs
$T$ for three underdoped samples with indicated
$T^*/T_c$. The arrows
show the corresponding values of $T_c$ which progressively decrease
 as $T^*/T_c$
increases. The inset plots (unpublished) experimental data on
 $\mathrm{La}_{2-x}\mathrm{Sr}_{x}\mathrm{Cu}\mathrm{O}_4$
from Ref.~\onlinecite{Ong3}
which curves are for $x=0.12$, $0.07$, and $0.03$ and which
 are in rough correspondence to the values of $T^*/T_c$ in the
  three theory plots. 
%Both the theoretical and experimental values of $\alpha_{xy}$ are
% normalized by 
%the value of $\alpha_{xy}$ for optimally doped sample at $2T_c$\cite{Huse}.
%$0.0232 (B/\mathrm{T})\mathrm{V}/(\mathrm{K\Omega m})$. This number was
%obtained earlier\cite{Huse} for $\alpha_{xy}$ in
%the optimal phase 
%when it is assumed that $\xi_{10}=30\mathrm{\AA}$
%and the c-axis inter-layer spacing is $6.615\mathrm{\AA}$.
%
%If we estimate
%the onset temperatures for $\alpha_{xy}$, called $T_{\alpha}^*$,
% using the experimental data to
%calibrate the fermionic background, we find $T^*/T_{\alpha}^*$ is given
%by 1.3, 1.7, and 2.6  for the solid, dashed and dotted curves, respectively.
The onset temperatures are of the order of $T^*/2$, as in experiment.
In the vicinity of $T_c$, the calculated behavior of $\alpha_{xy}$ corresponds
to that of TDGL theory, albeit with modified coefficients.  Thus $\alpha_{xy}$
diverges at $T_c$ although $\nu$ is finite there\cite{Huse}.
 Similarly, with overdoping
the behavior becomes characteristic of a more conventional fluctuation
picture\cite{Huse} in which $T^* \approx T_c$.  The essential distinction
between TDGL and the present case is that the onset temperature for bosonic
contributions can be substantially higher than the conventional fluctuation
regime. Semi-quantitative agreement between theory and experiment appears 
quite satisfactory.

In both theory and experiment
the dotted lines are for a non-superconductor with $T_c = 0$. 
To apply the present theory to the insulating case
we presume that the first term on the right hand side
of Eq.~\eqref{eq:a_1} is a non-vanishing number which corresponds
to the chemical potential of non-condensed electron pairs. This
constant is chosen to be 24K in order to fit the
maximum of $\alpha_{xy}(T)$ to its experimental counterpart.
This figure illustrates the
fact that $\alpha_{xy}$ (as well as $\nu$) vanishes at $T=0$ in
non-superconducting samples.
This derives from the behavior of the Bose function
which approaches a step function of $\omega$ at very low temperatures.

An interesting inference from the data\cite{Ong2} is that, while,
there is a considerable precursor effect for $\alpha_{xy}$,
the (orbital) magnetization drops to its superconducting
value only in the immediate vicinity of Bose condensation.  
This quantity
[given as $M^0_{ab}$ in Eq.~\eqref{eq:M}]
 is plotted in the inset to Fig.~\ref{fig:sigma},
for the same parameter set as in Fig.~\ref{fig:alpha}. 
The fermionic
background (measured experimentally) is neglible,
 so that the bosonic contribution
necessarily dominates.
The sharpness of the Meissner onset (which clearly reflects
$T_c$ and not $T^*$) can be attributed to the
small ratio of the boson velocity to the speed of light (and the small size of
the hyperfine constant). Similar results hold in TDGL-based calculations.

\subsection{AC Conductivity}

\begin{figure}[!thb]
\centerline{\includegraphics{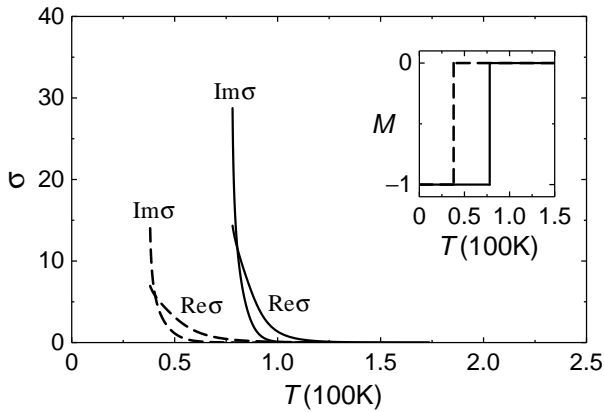}}
\caption{Real and imaginary components of 
normal state $\sigma$ (normalized
%\cite{Corson1999}
by $\sigma_Q\equiv e^2/s$) at $100~\mathrm{GHz}$ vs $T$. 
Solid and dashed curves have same
$T^*/T_c$ as for counterparts in Fig.~\ref{fig:alpha}.
 Associated magnetizations
are plotted in inset,
with the external magnetic field chosen as the unit.\label{fig:sigma}}
\end{figure}

In the main body of Fig.~\ref{fig:sigma}
are plotted the real and imaginary components of the
ac conductivity as a function of $T$ for $\omega/(2\pi) \approx 100~\mathrm{GHz}$ for
two different values of $T^*/T_c$, as in the previous figure.  
%In this subsection and the next, the conductivity $\sigma$ is normalized
%by $\sigma_Q\equiv e^2/s$, where $s$ is the inter-layer
%spacing; similarly, the resistivity is normalized by
%$\rho_Q\equiv 1/\sigma_Q$\cite{Corson1999}.
%
Both $\sigma_1 = \realpart \sigma$
and $\sigma_2 = \imaginarypart \sigma $
 are finite at $T_c$ with $\sigma_2$
larger than $\sigma_1$.  Because of the
relatively ``high" frequencies 
($\omega\approx 5\mathrm{K}\gg\widetilde{\Epsilon}_c$)\cite{Footnote_other},
the associated frequency dependence \textit{is not}
the asymptotically low $\omega$ limit, near $T_c$, where it would
vary
as $1/\sqrt{\omega}$.
%, because $\omega\approx
%5\mathrm{K}\gg\widetilde{\Epsilon_3}$.
It follows that, just as in TDGL theory, $\dif\sigma_1/\dif T$ is  finite
while $\dif\sigma_2/\dif T$ diverges at $T_c$.
We find 
that the magnitude of $\sigma$ is about twice the experimental value at $T_c$\cite{Corson1999}.
This prediction is the most notable difference between theory and
experiment\cite{Corson1999}. 
%It's possible that the microscopic inhomogeneity of the sample
%may reduce the ac conductivity near $T_c$. Recall that doped cuprates are not perfact
%periodic lattice, because of the doped atoms. This may affect the dynamics of the bosons.
%Another possibility is the critical behavior, which shows up in the critical regime.
%Nevertheless, the uncorrelated boson model on a periodic lattice
%does generate results resembling the experimental data when $T$ is not too close to $T_c$.

The theoretically deduced
onset temperature $T_{\sigma}^*$, for precursor effects in $\sigma_2$ 
can be estimated by noting that the fermionic background
 is relatively negligible
so that $\sigma_2$ becomes appreciable when it reaches
 a few percent
of its value at $T_c$.  This corresponds to onset temperatures
which are factors of 2 or so closer to $T_c$ than those estimated from $\alpha_{xy}$.

\begin{figure}[!thb]
\centerline{\includegraphics
{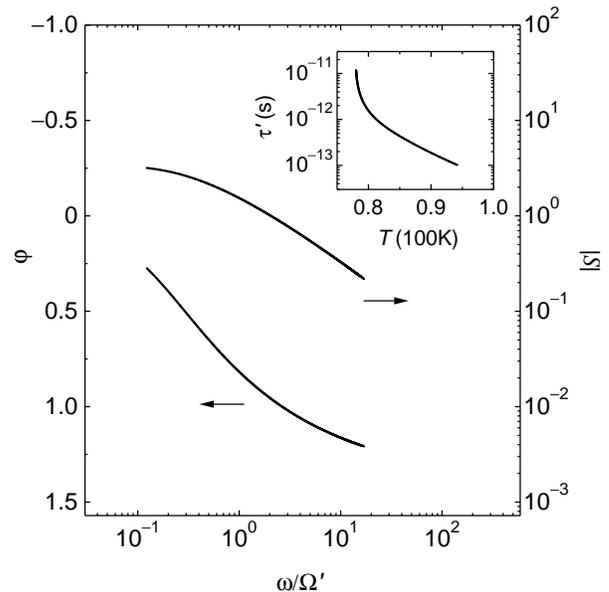}}
\caption{Fits to Kosterlitz-Thouless (KT) scaling of the 
conductivity which can be compared to
the analysis in Ref.\onlinecite{Corson1999}.
See text
for details.\label{fig:KT}} 
\end{figure}

In Fig.~\ref{fig:KT} we replot
the calculated ac conductivity
(for $T^*/T_c = 3.1$),
%\textbf{for the sample corresponding to the solid curves
%in Figs.~\ref{fig:alpha} and \ref{fig:sigma} },
following the Kosterlitz-Thouless (KT) -based
analysis of Ref.\onlinecite{Corson1999}. In the main
figure we illustrate these KT fits. These are the basis of
 an interpretation
of conductivity data in terms of vortices above $T_c$.
We present plots of
$\phi \equiv \tan^{-1} (\sigma_2/\sigma_1)$
and $\lvert S \rvert $ as a function of $\omega/\Omega'$
obtained from
%in the inset
%$1/\tau_{\sigma} \equiv \Omega'^{-1}$ as
%a function of $T$.  
%where both $\Omega$ and $1/\tau_{\sigma}$ are determined
$\sigma(\omega)/\sigma_Q= T_{\theta}^0 (T) S(\omega/\Omega') \Omega'^{-1}$. As in
the data, here,
$T_{\theta}^0$ and $\Omega'$ are deduced parameters.  The inset plots
$\tau' = 1/\Omega'$ as a function of temperature. We find that
$T_{\theta}^0$ is roughly constant in $T$, in contrast to the data which
finds this quantity to be decreasing as $T$ increases. Nevertheless,
 the agreement with experiment
for all three quantities plotted in Fig.~\ref{fig:KT} is quite good.
\textit{In this way one might argue that
key features of the ac conductivity data in Ref. \onlinecite{Corson1999}
and attributed to KT physics, may equally well
be explained by pre-formed
pairs}.

\subsection{Resistivity}

\begin{figure}[!thb]
\centerline{\includegraphics{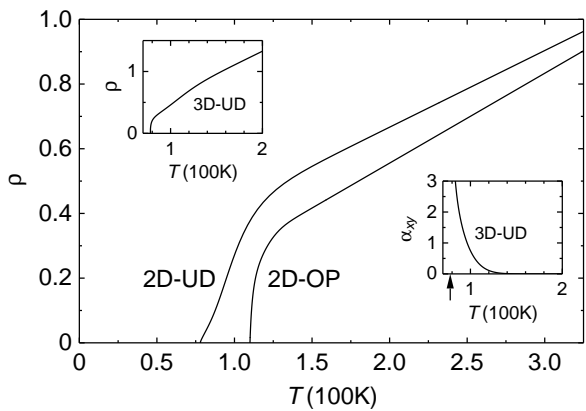}}
\caption{DC resistivity $\rho$ (normalized by $\rho_Q\equiv s/e^2$)
vs. $T$, for underdoped and optimally doped samples. The underdoped sample has the
same parameters as in previous figures. The optimally doped sample 
corresponds to
$\eta=1$. In the insets are plotted the $\rho$
and $\alpha_{xy}$
curves for an underdoped, more 3-dimensional,
system like YBCO, with $\eta=7.5$, and $T^*/T_c=3.1$. The arrow in the lower right inset indicates
$T_c$.
\label{fig:resistivity}}
\end{figure}

In the main body of Fig.~\ref{fig:resistivity} we plot DC resistivity curves,
$\rho$ vs. $T$ for an optimal
and underdoped system. In addition to the bosonic contribution
we have added a fairly generic\cite{Leridon2001PRL} linear-in-$T$
contribution from the fermions, as the background term.
The underdoped sample has the same parameters as in previous
figures with $T^*/T_c = 3.1$. For the optimally
doped case we take $T^*/T_c \approx 1.45$ and
$\eta \approx 1$.  It is clear that the transition
region is wider for the underdoped system. 
The increased separation between
$T^*$ and $T_c$, and the much
elongated boson relaxation time, are responsible for this widening.
Our optimal results compare favorably with those in the 
literature\cite{Balestrino1992PRB}.
For the underdoped case,
the wider than experimentally
observed\cite{Watanabe1997PRL}
transition regime that we find in $\rho$ is closely
connected to the behavior seen in the ac conductivity of
Figure 2. And the latter is compatible with the experiments
of Ref.\onlinecite{Corson1999}.

Dimensionality also plays an important role in determining
the behavior of the paraconductivity.
In the upper left inset of Fig.~\ref{fig:resistivity},
we plot the resistivity for a comparably underdoped, but
more three dimensional material, such as YBCO.
Increased 3-dimensionality (\textit{via} increased c-axis
coherence length) clearly shrinks the appararent transition range of $\rho$.
This 
can be traced to the reduced bosonic density of states in the low energy regime.
To see how the change of dimensionality
affects the Nernst coefficient, we
present a plot of $\alpha_{xy}$ in the lower inset for
this same sample. The onset for Nernst is
roughly 40K above $T_c$.  This contrasts sharply
with that of
$\rho$ which is at most a few K above $T_c$.
\textit{Thus, the same sample, with exactly
the same parameters, appears to have very different onset temperatures 
for resistivity and Nernst.}
It should be noted that if one looks closely at the plot of $\rho$
a small downward
curvature develops well above $T_c$, which can be
traced to $T^*$ effects.

\section{Conclusions}

In this paper we have investigated the effects of pre-formed
pairs or bosonic degrees of freedom on normal state 
transport properties. Our goal was to provide insight
into  
transport anomalies
in the high temperature superconductors. These anomalies are
associated with non-Fermi liquid-like observations, many of
which seem to be rather similar 
to what is found in the presence
of superconducting fluctuations.  By contrast to the standard
fluctuations of TDGL theory, however, these pre-formed pairs
appear at relatively high temperatures $T^*$ compared to
$T_c$. Throughout our discussion, it should be stressed that
we view $T^*$ as a ``crossover" temperature, rather than a sharp
phase transition. 
The origin of these preformed pairs is not specified. They
may arise from the strong
attractive interaction which drives the superconductivity.
(Indeed, attractive interactions in a \textit{non} $s$-wave
channel\cite{Dongzi} are possible,
despite the presence of strong Coulombic effects).
However, one may entertain as well other scenarios for the mechanism
of pair formation.

In the process of addressing transport we have devised a  
new formalism for extending
TDGL into the quantum regime, well away from
the transition temperature where
the bosons condense. Our approach, moreover, allows exact calculations
of all transport properties and, in the immediate
vicinity of $T_c$, our results are equivalent to those
of standard TDGL.
Comparison with experiment is quite satisfactory.
It should be stressed that no particular fitting to the data was done,
but rather in this paper we have explored the generic
features of the model. Moreover,
we have addressed a wide variety of different
experiments: Nernst, ac and dc conductivity and diamagnetic
measurements.
Aside from introducing a quantum statistical treatment of
the bosons, a key feature of our approach is the introduction
of the important temperature $T^*$.  This is a new concept, not
present in TDGL. Pairs now form
far from the critical regime. But at the same time
$T^*$ provides an ultimate cut-off.  There are no bosons
(or fermionic pseudogap) beyond this regime.

Our paper shows that transport anomalies are
compatible with the presence of
bosonic degrees of freedom. The same inference
is made from thermodynamic and ARPES experiments which deduce
the onset of a gap in the fermionic spectrum.  Indeed, we view
these broad classes of
phenomena as two sides of the same coin. One of the key
conclusions of this paper is that onset temperatures for
transport anomalies vary from one experiment to another.
Moreover, all
are considerably lower than $T^*$. The boson contribution
to transport depends sensitively on the
pair life time $\tau$.
Only sufficiently long lived bosons
significantly contribute to transport. Hence bosonic contributions
to transport are suppressed well before $T^*$ is reached.

Ours should be viewed as an alternative to the vortex scenario
or related phase fluctuation picture for addressing 
normal state transport anomalies.
Indeed, it has been recently suggested\cite{Lee} that phase
fluctuations alone cannot explain pseudogap phenomena. The
reasonable
agreement between the \textit{generic} results of the present
theory and experiment for all
the figures provides support
for a preformed-pair alternative to this vortex scenario.
Non-condensed bosons  
are also present below $T_c$ and, thereby, will lead to some continuity between
transport coefficients across the transition. In extending this work
to the \textit{ordered} phase, however, it will be useful to find
a relationship between bosons and vortices. Along these lines is
  the dual representation explored earlier\cite{Fisher}
in which vortices are the basic particles, rather than Cooper pairs.

Work supported by NSF-MRSEC Grant No.DMR-0213745. We thank
Y. Wang and N. P. Ong for unpublished data, and I. Ussishkin 
and G. Mazenko
for very useful conversations.

%\bibliography{Ref2}

\begin{thebibliography}
\bibitem[{foo({\natexlab{a}})}]{footnote2}
\bibinfo{note}{In the present case there is no particle-hole symmetry and we
  consider $T$ away from $T_c$ so that the thermal surface magnetization is
  non-zero.}
\bibitem{Footnote_gamma1}
If we are to study the Hall conductivity etc, the first order
term in $\gamma_1$ is important, since the zeroth order term vanishes. 
See Ref.\onlinecite{Huse}.
\bibitem{Footnote_widetilde}
Here the quantity that is determined is actually $\Epsilon_a/\gamma_2$. A
 similar situation holds
for $\Epsilon_c$ and $\mu_\text{pair}$.
Due to the approximate
linearity of $\widetilde{\Sigma}(\omega)$, 
only combinations such as $\mu_\text{pair}/\gamma_2$
or $m\gamma_2$ appear in low frequency ($\ll\Omega$) observables.
\bibitem{Footnote_n_Tc}
This formula assumes that $\gamma_2\Omega<\Epsilon_a$ and $\widetilde{\Epsilon}_c\ll T$.
For larger values of $\Omega$, the first term still increases with the ratio $\Omega/\Epsilon_a$,
but slower than $\Omega/(4\pi\Epsilon_a)$. For $\widetilde{\Epsilon}_c$ comparable to or
greater than $T$, the second term will decay slower with increasing $\widetilde{\Epsilon}_c$,
and won't be zero or even negative, as this small $\widetilde{\Epsilon}_c$ formula would suggest.
\bibitem{Footnote_transport}
The pair life time is thus on the order of $1/\widetilde{\Omega}$ at $(T_c+T^*)/2$.
If $1/\widetilde{\Omega}\ll 1/T$, as seems to be the case according to our analysis,
this implies that bosonic transport is negligible well before $T^*$ is reached,
as is confirmed by all transport experiments including the Nernst effect.
\bibitem{Footnote_other}
We write $T_c=\max(1.6T^*-0.5{T^*}^2-0.18,0)$ in units of roughly 100K, to semiquantitatively
fit the phase diagram of hole-doped cuprates. For all the 2D systems, we take $\widetilde{\Epsilon}_c
\approx 0.05$K, while for the 3D system illustrated in the insets of Fig.~\ref{fig:resistivity},
we take $\widetilde{\Epsilon}_c\approx 15$K.
 These are chosen
to be consistent with the observed 3D-2D crossover temperatures
of various materials.
\end{thebibliography}

\end{document}